\documentclass[11pt,a4paper]{article}
\usepackage[utf8]{inputenc}
\usepackage[T1]{fontenc}
\usepackage{lmodern}
\usepackage[english]{babel}
\usepackage[margin=2.5cm]{geometry} % Page margins
\usepackage{graphicx} % For including figures
\usepackage{booktabs} % For better-looking tables
\usepackage{natbib} % For author-year citations
\usepackage{amsmath} % For mathematical equations
\usepackage{amssymb} % For AMS symbols
\usepackage{setspace} % For line spacing
\usepackage{caption} % For customizing captions
\usepackage[colorlinks=true, citecolor=blue, linkcolor=blue, urlcolor=blue]{hyperref} % For hyperlinks
\usepackage{tabularx}
\usepackage{array}
\newcolumntype{R}[1]{>{\raggedright\arraybackslash\hsize=#1\hsize}X}

\usepackage{subcaption}
\setcitestyle{round}

\let\oldfootnote\footnote
\renewcommand{\footnote}{\fontsize{9}{11}\selectfont\oldfootnote}

\title{StatCite: A Large-scale Citation Network Dataset for Statistics and Data Science}
\author{
    Tianang Deng \\
    Central University of Finance and Economics \\
    \and 
    Tianchen Gao \\
    Peking University \\
    \and 
    Rui Pan \thanks{Corresponding author: ruipan@cufe.edu.cn} \\
    Central University of Finance and Economics \\
    \and
    Yan Zhang \\
    Shanghai University of International Business and Economics
}
\date{} % Remove date

\begin{document}

\maketitle

\begin{abstract}
In this paper, we introduce \textbf{StatCite}, a large-scale citation network dataset covering publications in statistics and data science from 1981 to 2025. The dataset contains 189,101 research articles collected from 62 representative journals and provides bibliographic metadata, including title, author list, publisher, published year, abstract, keywords,  and reference list.
Based on the collected publications, we construct four complementary citation-based networks, namely the paper citation network, the co-citation network, the bibliographic coupling network, and the journal citation network. To illustrate the utility of the dataset, we present descriptive analyses of the constructed networks and investigate the community structure of the paper citation network. The results show that StatCite preserves key structural characteristics commonly observed in large-scale citation networks and captures several major research areas in statistics and data science. By integrating multiple network representations with rich textual metadata, StatCite provides a valuable resource for statistical analysis, knowledge discovery, and data-driven studies of scientific literature.
\end{abstract}

% AMS subject classification

\textbf{Keywords:} Bibliometrics, Citation Networks, Multi-layer Networks, Statistics and Data Science

\textbf{Mathematics Subject Classification (2020):} 62R10

\section{Introduction}

As the volume of scientific publications increases steadily, the relationships among papers, authors, and journals become more complex. Understanding how knowledge emerges, evolves, and interacts across scientific domains has become a central topic in quantitative science studies \citep{fortunato2018science}. Citation networks provide a powerful framework for this purpose. In a citation network, the nodes represent publications, authors, or journals, and the edges represent the citation links that connect them \citep{gao2023large}. This representation allows researchers to model the flow of scientific information, identify communities of related research, and measure patterns of influence over time \citep{mingers2015review,mejia2021exploring}. Consequently, citation networks have been extensively applied in disciplines such as physics \citep{teich2022citation}, digital marketing \citep{krishen2021broad}, biology \citep{feng2024citation}, statistics \citep{gao2021community,gao2024community,liu2025academic} and others. These studies show that network-based approaches provide valuable insights into the structure and dynamics of science. 

Citation networks and their related structure allow for several types of analyses. At the paper level, a citation network records links from a given paper to earlier work it cites, which reveals how new work builds on older ideas \citep{price1965networks}. A co-citation network links two papers when they are both cited by the same later work, thus capturing intellectual coupling or topical similarity \citep{chen2010structure}. Bibliographic coupling network arises when two papers cite the same previous work and thus offers another perspective on how knowledge is shared \citep{kessler1963bibliographic}. At a higher level, journal citation networks enable the study of how different venues relate to one another and how research influence circulates within and across disciplines \citep{varin2016statistical}. However, to conduct these analyses appropriately, we need datasets with clean metadata, reliable reference linking, and consistent author and journal information \citep{bornmann2025citation}.

Large citation datasets already exist for some fields. For instance, high-energy physics has the cit-HepPh dataset, covering 34,546 papers and 421,578 citations from arXiv repository \citep{leskovec2005graphs}. Citation graphs derived from databases such as DBLP or ACM/Google Scholar include millions of papers and extensive citation connections across sub-disciplines. In life sciences and biomedical research, open citation resources, such as the NIH Open Citation Collection, make available large numbers of link-level citation records between publications indexed in PubMed \citep{hutchins2019nih}. Despite these advances, fields such as statistics, econometrics and data mining remain under-served by domain-specific large-scale citation datasets. The team of \cite{ji2016coauthorship} construct one of the first cleaned co-authorship and citation networks for statisticians, and \cite{ji2022co} extend this work in scale and temporal span. Recently, \cite{He12012026} collect a dataset of papers from four top-tier statistical journals and three conferences. Based on this dataset, they propose a topic model and examine the evolution of research topics over the past four decades. Their efforts demonstrate that methodological fields can benefit from network-based bibliometric data. 

In this work, we focus on the construction and release of a large-scale citation dataset, i.e., {\bf StatCite}. Specifically, we collect 189,101 publications from 62 journals in statistics, econometrics, and data mining, covering the period from 1981 to 2025. We further construct four types of citation-based networks, including the paper citation network, the co-citation network, the bibliographic coupling network, and the journal citation network. These network representations provide a unified framework for analyzing scholarly knowledge at multiple levels. Table~\ref{tab:comparison} summarizes the main differences between our dataset and two representative citation datasets in statistics-related fields \citep{ji2022co, He12012026}.
To illustrate the usefulness of the dataset, we present descriptive analyses of the networks and examine their structural properties, as well as the community structure identified in the paper citation network. The dataset and the corresponding network constructions are made publicly available to support future research on methodological development (\url{https://github.com/Gaotianchen97/Dataset-StatCite}).

\begin{table}[htbp]
    \centering
    \small % 稍微缩小字号以容纳更多内容
    \renewcommand{\arraystretch}{1.35} % 增加行距，提高可读性
    \caption{Comparison of Datasets}
    \label{tab:comparison}
    % 定义表格列格式
    % p{0.18\textwidth}: 第一列固定宽度
    % 后三列分配剩余空间，并自动换行
   \begin{tabularx}{\textwidth}{@{}
  >{\raggedright\arraybackslash\hsize=0.6\hsize}X
  >{\raggedright\arraybackslash\hsize=1.1\hsize}X
  >{\raggedright\arraybackslash\hsize=1.1\hsize}X
  >{\raggedright\arraybackslash\hsize=1.2\hsize}X @{}}
        \toprule
        \textbf{Data} & \textbf{\cite{ji2022co}} & \textbf{ \cite{He12012026}} & \textbf{Our Data} \\
        \midrule
        
        \textbf{Time Span} 
        & 1975 -- 2015 
        & 1980 -- 2024 (Journals) \newline 2014 -- 2024 (Conferences) 
        & 1981 -- 2025 \\ 
        \midrule
        
        \textbf{Sample Size} 
        & 83,331 papers \newline 47,311 authors 
        & 47,530 raw papers \newline (27,735 after pre-selection filtering) 
        & 189,101 papers \\ 
        \midrule
        
        \textbf{Source} 
        & 36 journals including AOAS, Biostatistics, JMLR, Bernoulli, etc. 
        & JASA, AOS, JRSSB, Biometrika, NeurIPS, ICML, AISTATS 
        & 62 journals including JASA, AOS, JRSSB, Biometrika, JOE, TKDE, etc. \\ 
        \midrule
        
        %\textbf{Has Paper ID?} & Yes & Yes & Yes \\ 
        %\midrule
        
        %\textbf{Has Author ID?} & Yes & No & {\color{red} Yes} \\ 
        %\midrule
        
        \textbf{Metadata Fields} 
        & DOI, WoS ID, Title, Authors, Year, Journal, Volume, Issue, Pages 
        & Title, Authors, Year, Source, Abstract, Institution 
        & Title, Authors, Journal, Year, Abstract, Keywords, Reference List \\ 
        \midrule
        
        \textbf{Ready-to-Use Data} 
        & \textbf{AuPapMat:} Author \& Paper IDs, Year, Journal \newline
          \textbf{PapPapMat:} Citing \& Cited Paper IDs, Years, Self-cite indicator \newline
          \textbf{Network:} \newline
          -- Citee networks \newline
          -- Co-authorship networks 
        & \textbf{Processed Text Corpus:} \newline
        -- Phrase-segmented abstracts \newline
        -- BERT-filtered conference subset \newline
        -- Journal/conference category labels 
        & \textbf{Network:}  \newline
        -- Paper citation network \newline
        -- Co-citation network \newline
        -- Bibliographic coupling network \newline
        -- Journal citation network\\ 
        \midrule
        
        \textbf{Publicly Available} & Yes & Yes & Yes \\ 
        \bottomrule
    \end{tabularx}
\end{table}

The rest of this work is organized as 
follows. Section 2 introduces the data collection procedure and the data cleaning steps. In Section 3, four citation networks are constructed and summarized. Data usage is also presented in this section. Descriptive analysis as well as community detection are shown in Section 4. At last, Section 5 concludes the paper with potential use of our dataset.

\section{Data Collection and Processing}

This section describes the data sources, collection procedures, and processing steps used to construct the dataset. We collect publication records from the Web of Science (WoS) and complement them with additional data from AMiner to improve coverage in recent years. The raw data include bibliographic metadata and reference lists, which form the basis for constructing multiple citation-based networks. To ensure the reliability and usability of the dataset, we implement a series of data cleaning procedures, including duplicate removal, filtering of non-research items, and consistency checks across records. These steps aim to produce a coherent and structured dataset that can support subsequent network construction and empirical analysis.

\subsection{Data Collection}

\begin{figure}[!ht]
    \centering
    \includegraphics[width=0.9\linewidth]{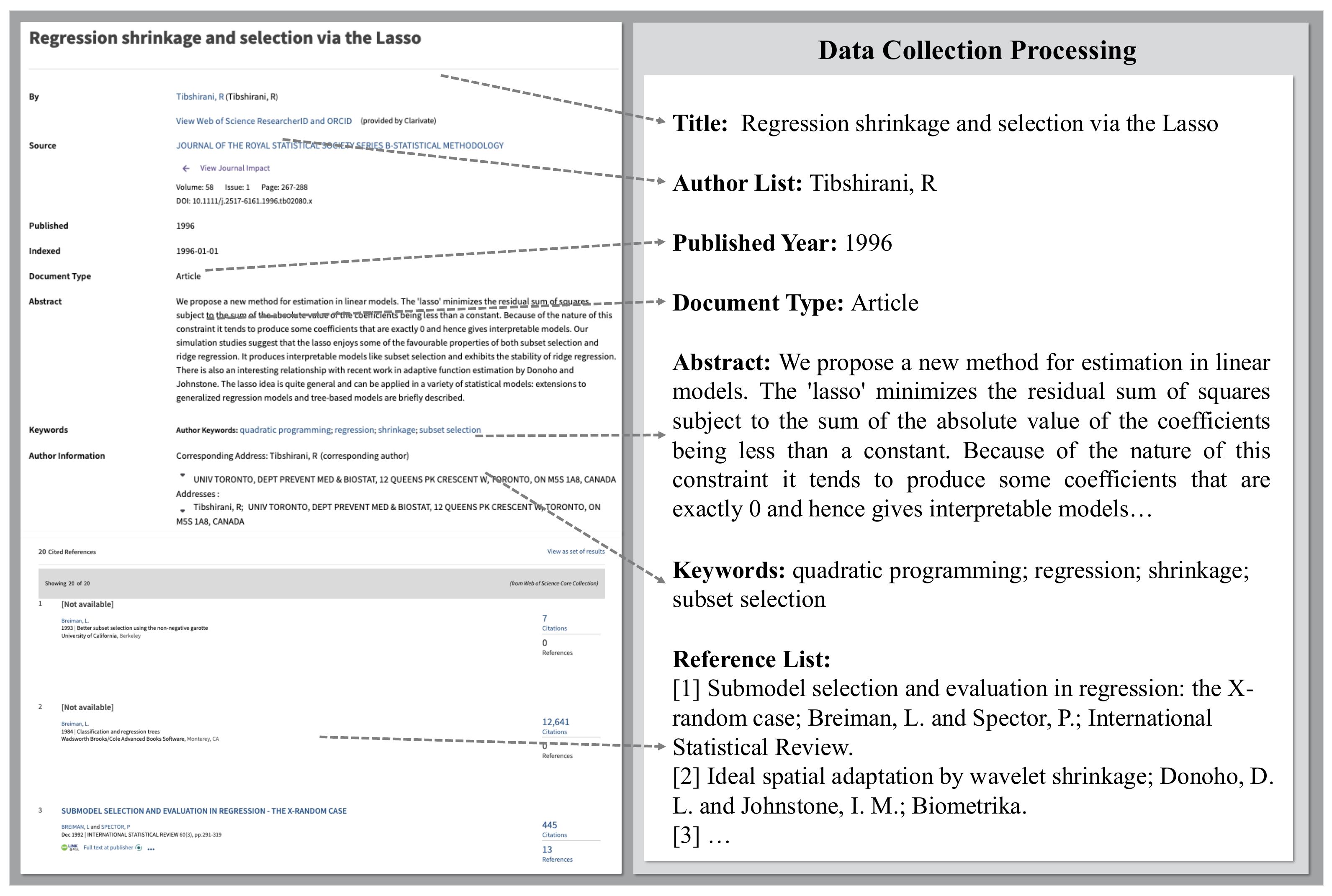}
    \caption{Illustration of the web crawling process and extracted paper metadata from WoS.}
    \label{fig:datacollection}
\end{figure}

We collect data from the WoS (\url{https://www.webofscience.com}) and AMiner (\url{https://www.aminer.cn/}). Our dataset covers 62 journals spanning statistics, econometrics, and computer science. The full list of journals is reported in Table~\ref{Tab:journals}. The journal selection is based on JCR quartiles, impact factors, and subject classifications. In particular, we focus on JCR subject areas of \emph{Statistics \& Probability}, \emph{Economics}, and \emph{Computer Science}. We implement web crawling procedures with parallel processing to collect paper metadata from WoS for the period 1981--2024. To complement these records, we retrieve additional data from AMiner via its API. This source provides papers published in late 2024 and 2025, as well as some earlier records not indexed in WoS. The collected information includes title, author list, publisher (i.e., journal), published year, document type (not available in the Aminer dataset), abstract, keywords, and reference list. Figure~\ref{fig:datacollection} illustrates the data collection process from WoS, showing how paper metadata are extracted from web pages.

\begin{table}[!htbp]
    \tiny
    \caption{The 62 selected journals and summary statistics of collected papers, listed in alphabetical order.}
    \label{Tab:journals}
    \centering
    \renewcommand{\arraystretch}{1.15}
    \setlength{\tabcolsep}{4pt}

    \begin{tabular}{
        c
        >{\raggedright\arraybackslash}p{7.2cm}
        c
        c
        c
    }
        \hline
        \hline
        \textbf{ID} &
        \textbf{Journal} &
        \textbf{Abbreviation} &
        \textbf{Period} &
        \textbf{\# of articles} \\
        \hline

        1  & ACM Transactions on Knowledge Discovery from Data
           & -- & 2007--2025 & 1,263 \\

        2  & Advances in Data Analysis and Classification
           & -- & 2007--2025 & 502 \\

        3  & American Statistician
           & -- & 1981--2025 & 2,093 \\

        4  & Annals of Applied Statistics
           & AOAS & 2007--2025 & 1,776 \\

        5  & Annals of Statistics
           & AOS & 1981--2025 & 4,418 \\

        6  & Annals of the Institute of Statistical Mathematics
           & -- & 1981--2025 & 2,075 \\

        7  & Annual Review of Statistics and Its Application
           & -- & 2014--2025 & 221 \\

        8  & Australian \& New Zealand Journal of Statistics
           & -- & 1998--2025 & 846 \\

        9  & Bayesian Analysis
           & -- & 2006--2025 & 797 \\

        10 & Bernoulli
           & -- & 1995--2025 & 2,084 \\

        11 & Bioinformatics
           & -- & 1985--2025 & 18,550 \\

        12 & Biometrics
           & Bcs & 1981--2025 & 5,474 \\

        13 & Biometrika
           & Bka & 1981--2025 & 3,634 \\

        14 & Biostatistics
           & -- & 2000--2025 & 1,366 \\

        15 & Canadian Journal of Statistics
           & -- & 1981--2025 & 1,451 \\

        16 & Communications in Statistics-Simulation and Computation
           & -- & 1981--2025 & 6,076 \\

        17 & Communications in Statistics-Theory and Methods
           & CSTM & 1983--2025 & 11,973 \\

        18 & Computational Statistics
           & -- & 1999--2025 & 1,885 \\

        19 & Computational Statistics \& Data Analysis
           & CSDA & 1983--2025 & 6,277 \\

        20 & Data Mining and Knowledge Discovery
           & -- & 1997--2025 & 1,201 \\

        21 & Econometrica
           & ECTA & 1981--2025 & 2,838 \\

        22 & Electronic Journal of Statistics
           & -- & 2007--2025 & 2,036 \\

        23 & IEEE Transactions on Knowledge and Data Engineering
           & -- & 1989--2025 & 6,609 \\

        24 & International Statistical Review
           & -- & 1981--2025 & 948 \\

        25 & Journal of Applied Statistics
           & -- & 1981--2025 & 4,305 \\

        26 & Journal of Business \& Economic Statistics
           & -- & 1983--2025 & 2,103 \\

        27 & Journal of Classification
           & -- & 1984--2025 & 702 \\

        28 & Journal of Computational and Graphical Statistics
           & JCGS & 1992--2025 & 2,063 \\

        29 & Journal of Econometrics
           & JOE & 1981--2025 & 4,648 \\

        30 & Journal of Machine Learning Research
           & -- & 2000--2025 & 3,358 \\

        31 & Journal of Multivariate Analysis
           & JMVA & 1981--2025 & 4,491 \\

        32 & Journal of Nonparametric Statistics
           & -- & 1991--2025 & 1,397 \\

        33 & Journal of Statistical Computation and Simulation
           & -- & 1981--2025 & 4,150 \\

        34 & Journal of Statistical Planning and Inference
           & JSPI & 1981--2025 & 6,787 \\

        35 & Journal of Statistical Software
           & -- & 1997--2025 & 1,363 \\

        36 & Journal of Survey Statistics and Methodology
           & -- & 2013--2025 & 448 \\

        37 & Journal of the American Statistical Association
           & JASA & 1981--2025 & 6,048 \\

        38 & Journal of the Royal Statistical Society Series A-Statistics in Society
           & -- & 1981--2025 & 1,668 \\

        39 & Journal of the Royal Statistical Society Series B-Statistical Methodology
           & JRSS-B & 1981--2025 & 1,573 \\

        40 & Journal of the Royal Statistical Society Series C-Applied Statistics
           & -- & 1982--2025 & 1,805 \\

        41 & Journal of Time Series Analysis
           & -- & 1981--2023 & 1,757 \\

        42 & Machine Learning
           & -- & 1986--2025 & 2,495 \\

        43 & Neural Networks
           & -- & 1988--2025 & 6,978 \\

        44 & R Journal
           & -- & 2009--2025 & 749 \\

        45 & Review of Economics and Statistics
           & -- & 1981--2025 & 3,246 \\

        46 & Scandinavian Journal of Statistics
           & SJS & 1981--2025 & 1,870 \\

        47 & Spatial Statistics
           & -- & 2012--2025 & 779 \\

        48 & Stat
           & -- & 2012--2025 & 813 \\

        49 & Stata Journal
           & -- & 2001--2025 & 976 \\

        50 & Statistica Sinica
           & StatSin & 1991--2025 & 2,645 \\

        51 & Statistical Analysis and Data Mining
           & -- & 2008--2025 & 630 \\

        52 & Statistical Methods and Applications
           & -- & 2001--2025 & 809 \\

        53 & Statistical Methods in Medical Research
           & -- & 1992--2025 & 2,323 \\

        54 & Statistical Modelling
           & -- & 2001--2025 & 575 \\

        55 & Statistical Papers
           & -- & 1988--2025 & 2,150 \\

        56 & Statistical Science
           & -- & 1986--2023 & 825 \\

        57 & Statistics
           & -- & 1985--2025 & 2,017 \\

        58 & Statistics \& Probability Letters
           & SPL & 1982--2025 & 9,449 \\

        59 & Statistics and Computing
           & SC & 1991--2025 & 2,143 \\

        60 & Statistics in Medicine
           & SIM & 1982--2025 & 9,993 \\

        61 & Technometrics
           & Technom. & 1981--2025 & 1,639 \\

        62 & Test
           & -- & 1992--2025 & 938 \\

        \hline
    \end{tabular}
\end{table}

\subsection{Data Processing}

%We perform several data cleaning and preprocessing steps. First, we remove duplicate records and papers with missing key information. Second, we exclude non-research items such as discussions, comments, and other informal publications. Third, we address title ambiguity. We note that paper titles alone cannot uniquely identify articles, as different papers may share the same title. Therefore, we use a combination of title, journal name, and author list as a unique identifier for each paper. After these procedures, we identify 189,101 unique papers. The annual number of papers is shown in Figure~\ref{fig:numofpapers}.

We perform several data cleaning and preprocessing steps to ensure the consistency and reliability of the dataset. 
First, we remove duplicate records and papers with missing key information, such as titles or reference lists. To address title ambiguity and record inconsistency across data sources, duplicate detection is based on a combination of title, journal name, and author list, rather than title alone, which allows us to distinguish articles more accurately and avoid repeated inclusion of the same publication. 
Second, we exclude non-research items such as discussions, comments, editorial materials, and other informal publications. These records typically lack complete reference information and may introduce noise into the citation network, and are therefore removed to retain standard research articles. Specifically, for the records collected from WoS, we filter publications based on the document type field and retain only items classified as Article. For the records obtained from AMiner, where a comparable document type field is not available, we manually screen the retrieved records and exclude non-research materials to ensure consistency with the WoS-based dataset.
Third, we reconcile publication year information across data sources. In WoS, the recorded year may occasionally reflect the indexing year rather than the actual published year, causing differences of one to two years. Manual inspection shows that AMiner provides more accurate publication years in most cases, so we use AMiner as the primary source of year information. For a small number of papers (only about 100), however, AMiner may contain substantial year errors. When the difference between the two sources exceeds ten years, the WoS year is usually more accurate, and we therefore use it instead.
Fourth, we perform reference matching and citation filtering for network construction. Each article may cite numerous publications that are not included in the final paper collection. Since these external publications cannot be linked to corresponding paper records within our dataset, they are excluded when constructing citation-based networks. Specifically, we match the reference information of each paper against the final set of collected publications and retain only successfully matched citation relationships. Unmatched references are not used in the construction of citation, bibliographic coupling, and co-citation networks. The original reference information, including reference titles, is preserved in the released dataset. 
Overall, these preprocessing steps reduce data redundancy and improve the consistency of bibliographic and citation information. After these procedures, we identify 189,101 unique papers.

\noindent 
{\bf Remark 1.} The author list information is retained but not further processed in this study. Constructing co-authorship networks requires additional procedures to resolve author identity issues, including name ambiguity and variant author representations. Since the present study focuses on citation-based networks and does not construct co-authorship networks, we leave author normalization beyond the scope of this work. Nevertheless, the author information is preserved to facilitate future research on collaboration patterns.

\section{Network Construction}

Based on the cleaned bibliographic data, we construct several types of citation-based networks to represent scholarly relationships at different levels. These networks provide complementary views of knowledge flow and similarity among publications.

\subsection{Paper Citation Network}

A paper citation network provides the most fundamental representation of scholarly knowledge flows, where nodes correspond to individual publications, and directed edges represent citation relationships. Formally, a citation network can be represented by an adjacency matrix $A\in\mathbb{R}^{n\times n}$, where $A_{i_1i_2}=1$ if article $i_1$ cites article $i_2$ and $A_{i_1i_2}=0$ otherwise, with $n$ denoting the total number of articles. By definition, self-citations are excluded, and we let $A_{ii}=0$ for all $i = 1,\cdots, n$. In addition, because articles can only cite previously published work, the resulting citation network is a directed acyclic graph (DAG). This directed acyclic structure captures the directional accumulation of scientific knowledge and forms the primary data foundation for citation-based impact analysis. Nevertheless, while direct citation links reflect explicit referencing behavior, they do not fully characterize higher-order similarities among articles. This motivates us to construct the following derived citation-based networks.

\subsection{Derived Paper-Level Networks}

One such derived structure is the co-citation network, which emphasized how papers are jointly cited by subsequent works. Intuitively, two papers are considered related if they are frequently cited together, even if they do not directly cite each other. Mathematically, the co-citation network is an undirected weighted network represented by the adjacency matrix $C= A^\top A\in\mathbb{R}^{n\times n}$. The element $C_{i_1i_2} =\sum_{k=1}^n A_{ki_1}A_{ki_2}$ measures the number of papers that cite both $i_1$ and $i_2$. Larger values of $C_{i_1i_2}$ indicate stronger perceived similarity as recognized by the citing community. Unlike the original citation network, co-citation links are symmetric, reflecting how earlier papers are collectively positioned within the evolving literature. In contrast to co-citation, the bibliographic coupling network focuses on similarity from the perspective of shared references. Two papers are bibliographically coupled if they cite common prior work. Formally, the bibliographic coupling network is defined by the adjacency matrix $B=AA^\top\in\mathbb{R}^{n\times n}$, where $B_{i_1i_2}=\sum_{k=1}^n A_{i_1k}A_{i_2k}$ counts the number of references shared by papers $i_1$ and $i_2$. This undirected weighted network captures topical or methodological proximity at the time of publication and is therefore forward-looking relative to co-citation. Together, co-citation and bibliographic coupling provide complementary views of scholarly similarity.

\subsection{Journal Citation Network}

While the above networks operate at the paper level, aggregation over journals leads naturally to the journal citation network. This kind of network characterizes citation flows between journals. Specifically, the journal citation network is a directed and weighted graph with adjacency matrix $W=(W_{j_1j_2})\in\mathbb{R}^{m\times m}$, where
$W_{j_1j_2}$ represents the total number of citations from papers published in journal $j_1$ to those in journal $j_2$. This higher-level network smooths individual citation noise and highlights disciplinary interactions and journal influence structures. To summarize, paper-level and journal-level citation networks form a coherent multi-scale framework for analyzing the organization and evolution of scientific knowledge. 

\subsection{Summary of Networks}

Table \ref{tab:overview} summarizes the main characteristics of the constructed citation-based networks. Specifically, the paper citation network is a directed graph with more than 150,000 nodes and over one million edges, capturing explicit citation links among publications. In addition, the co-citation and bibliographic coupling networks are undirected and weighted, with substantially more edges due to connections formed through shared citation patterns. Co-citation reflects how papers are cited together, while bibliographic coupling captures shared references. At a higher level, the journal citation network aggregates citation relationships across 62 journals. Compared with paper-level networks, it is much smaller but denser, and highlights citation flows between research venues. These networks together provide a multi-scale representation of the dataset and support analysis at both the paper and journal levels.

\begin{table}[htbp]
\centering
\small % 稍微缩小字号以容纳更多内容
\renewcommand{\arraystretch}{1.4} % 增加行距，提高可读性
\caption{Overview of the constructed citation-based networks, including network size, edge definitions, and structural types. The reported numbers of nodes correspond to papers that participate in at least one edge in the corresponding network. Isolated papers without any citation, co-citation, or bibliographic coupling relationships within the collected dataset are excluded from the network construction.}
\label{tab:overview}
\begin{center}
\resizebox{\textwidth}{!}{
\begin{tabular}{cccccc}
\toprule
\textbf{Network} & \textbf{Node Type} & \textbf{\# of Nodes} & \textbf{Edge Definition} & \textbf{\# of Edges} & \textbf{Type} \\
\midrule
Paper citation & Paper &  157,277 & Citation relationship & 1,022,238 & Directed, Temporal\\
Co-citation & Paper & 111,303 & Co-citation relationship & 4,431,138 & Undirected, Weighted, Temporal\\
Bibliographic coupling & Paper & 131,914 & Shared
reference & 42,198,754 & Undirected, Weighted \\
Journal citation & Journal & 62 & Journal-to-journal citation & 3,700 & Undirected, Weighted, Temporal\\
\bottomrule
\end{tabular}
}
\end{center}
\end{table}

\subsection{Data Usage}

The complete dataset can be downloaded from the accompanying GitHub repository at \url{https://github.com/Gaotianchen97/Dataset-StatCite}. The current release comprises eight logical datasets, each provided in both Apache Parquet and CSV formats. All textual fields are encoded in UTF-8. Parquet is recommended for most analyses because it provides efficient compression and supports selective reading, which is particularly useful for the large network tables.
The \textit{paper\_info} table contains metadata for 189,101 papers. Its fields include the internal paper identifier, title, author list, publisher or publication venue, publication year, abstract, keywords, and the complete reference list associated with each paper. The \textit{reference\_list} field stores the full bibliographic references cited by each paper as a serialized list, rather than providing only the internal identifiers of references that can be matched within the dataset. The \textit{keywords} field is likewise stored as a serialized list.

Four aggregated network tables are provided. The \textit{citation\_network\_edgelist} table represents the directed paper citation network, in which an edge runs from \textit{citing\_paper} to \textit{cited\_paper}. The \textit{bib\_coupling\_network\_edgelist} table represents the undirected bibliographic-coupling network, where the edge weight indicates the number of references shared by two papers. The \textit{cocitation\_network\_edgelist} table represents the undirected co-citation network, where the edge weight indicates the number of times two papers are cited together. The \textit{journal\_citation \_network\_edgelist} table represents the directed journal citation network, with edges running from \textit{citing\_journal} to \textit{cited\_journal} and weights indicating citation frequency. Time-resolved versions are additionally provided for the paper citation, co-citation, and journal citation networks: \textit{citation\_network\_edgelist\_withyear}, \textit{cocitation\_network\_edgelist\_withyear}, and \textit{journal\_citation\_network\_edgelist\_withyear}. Each of these tables contains only one \textit{year} field. In the paper and journal citation networks, this field indicates the publication year of the citing paper. In the co-citation network, it indicates the year in which the co-citation occurred. Therefore, the year field should not be interpreted as the publication year of both nodes connected by an edge.

Users who require the publication years of individual papers can obtain them by matching \textit{citing\_paper}, \textit{cited\_paper}, \textit{Paper1}, or \textit{Paper2} to \textit{paper\_ID} in the \textit{paper\_info} table and retrieving the corresponding \textit{published\_year}. For the bibliographic-coupling and co-citation networks, the publication years of the two connected papers can be added separately through two matching operations. Other temporal variables or customized year combinations can be constructed in the same manner.
The \textit{paper\_ID} field is the primary identifier linking the paper metadata to the paper-level network tables. These identifiers are internal dataset identifiers and should be treated as character strings rather than numeric values. The network tables without the \textit{\_withyear} suffix aggregate relationships across all available years, whereas the corresponding time-resolved tables retain the year associated with each citation or co-citation event.

\section{Illustrative Data Analysis}

In this section, we present several analyses to illustrate the characteristics and potential usage of the constructed citation networks. We first examine the temporal evolution of publications and citation patterns, followed by an analysis of the structural properties of the derived networks. We then investigate the community structure of the paper citation network to show how the dataset can support downstream scientometric analysis.

\subsection{Temporal Evolution}

\begin{figure}[!ht]
    \centering
    \includegraphics[width=0.9\linewidth]{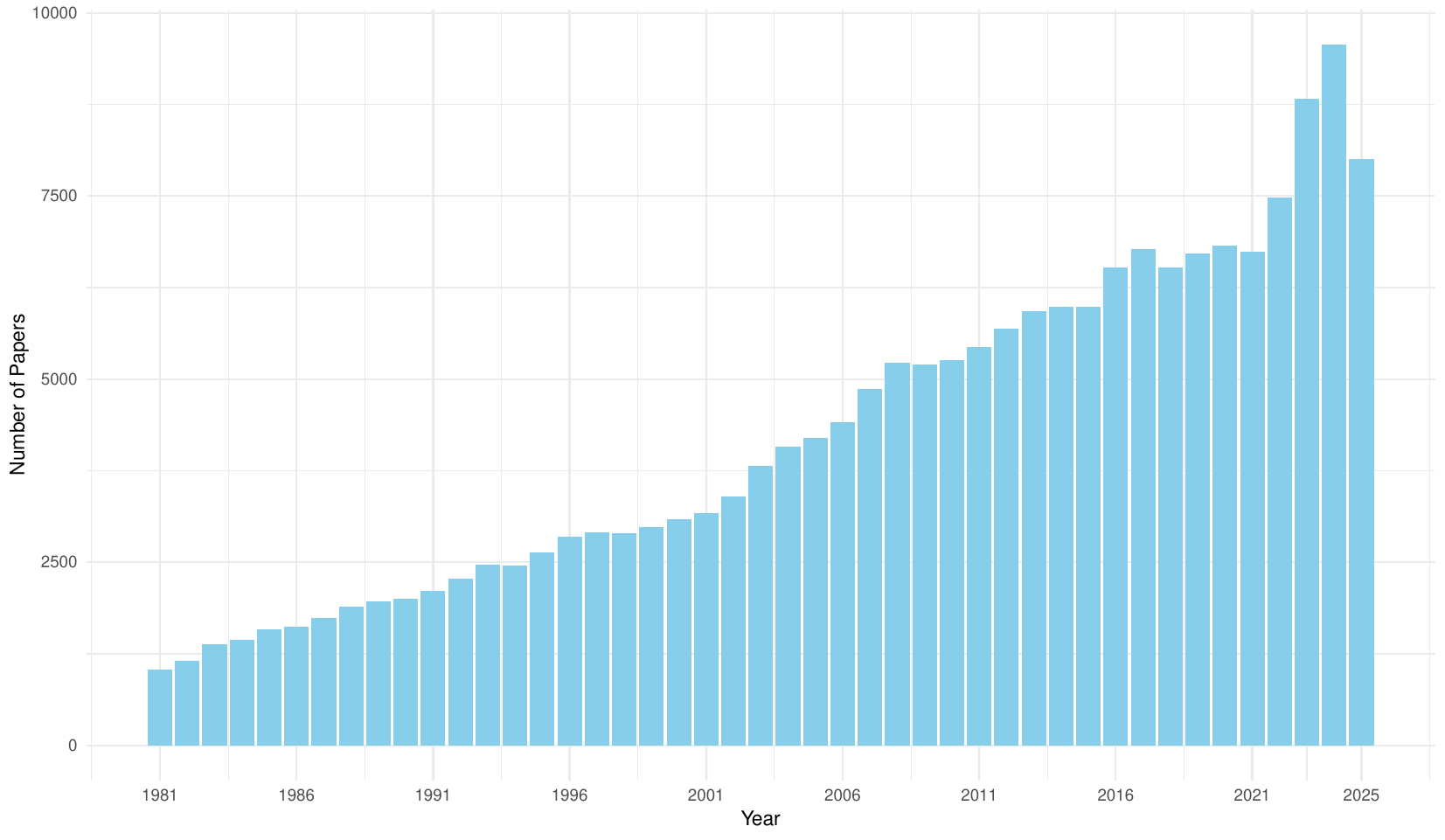}
    \caption{The annual number of papers in our dataset.}
    \label{fig:numofpapers}
\end{figure}

We begin by examining the temporal evolution of the dataset. Figure~\ref{fig:numofpapers} shows the annual number of publications from 1981 to 2025. The number of papers increases steadily throughout the study period, with noticeably faster growth after the early 2000s. This trend reflects the continuous expansion of research activity in statistics, econometrics, data mining, and related methodological disciplines. The growth in publications leads naturally to the expansion of the citation network. As new papers continuously enter the literature and establish citation links with previous work, both the number of nodes and citation relationships increase substantially over time. To further examine the structural evolution of the network, we compare its density at the beginning and end of the study period. Although both the number of papers and the number of citation links increase over time, the network density decreases from 0.1072\% in 1981 to 0.0031\% in 2025. This result indicates that the increase in citation links is relatively small compared with the growth of the network. Consequently, the citation network becomes increasingly sparse as the literature grows. The observed trend is consistent with findings from previous empirical studies of citation networks \citep{leskovec2005graphs} and suggests that the constructed dataset preserves important structural characteristics of scientific citation systems.

\noindent
{\bf Remark 2.} The relatively small number of publications in 2025 does not necessarily indicate a decline in research output. Recent publications may not yet be fully indexed in bibliographic databases due to publication and indexing delays. In addition, the availability of records may depend on the accessibility of data sources during the collection process. Therefore, the number of papers in the most recent year should be interpreted with caution, and future updates of the dataset may further improve the coverage of recent publications. In addition, the reference information for papers published in 2025 is currently incomplete in the AMiner database. As a result, reference lists are missing for a large proportion of the 2025 publications included in this dataset, limiting their participation in the construction of citation-based networks. These missing reference data will be incorporated in future release.

\subsection{Citation Patterns}

\begin{figure}[htbp]
    \centering
    \begin{subfigure}[b]{0.48\textwidth}
        \centering
        \includegraphics[width=\textwidth]{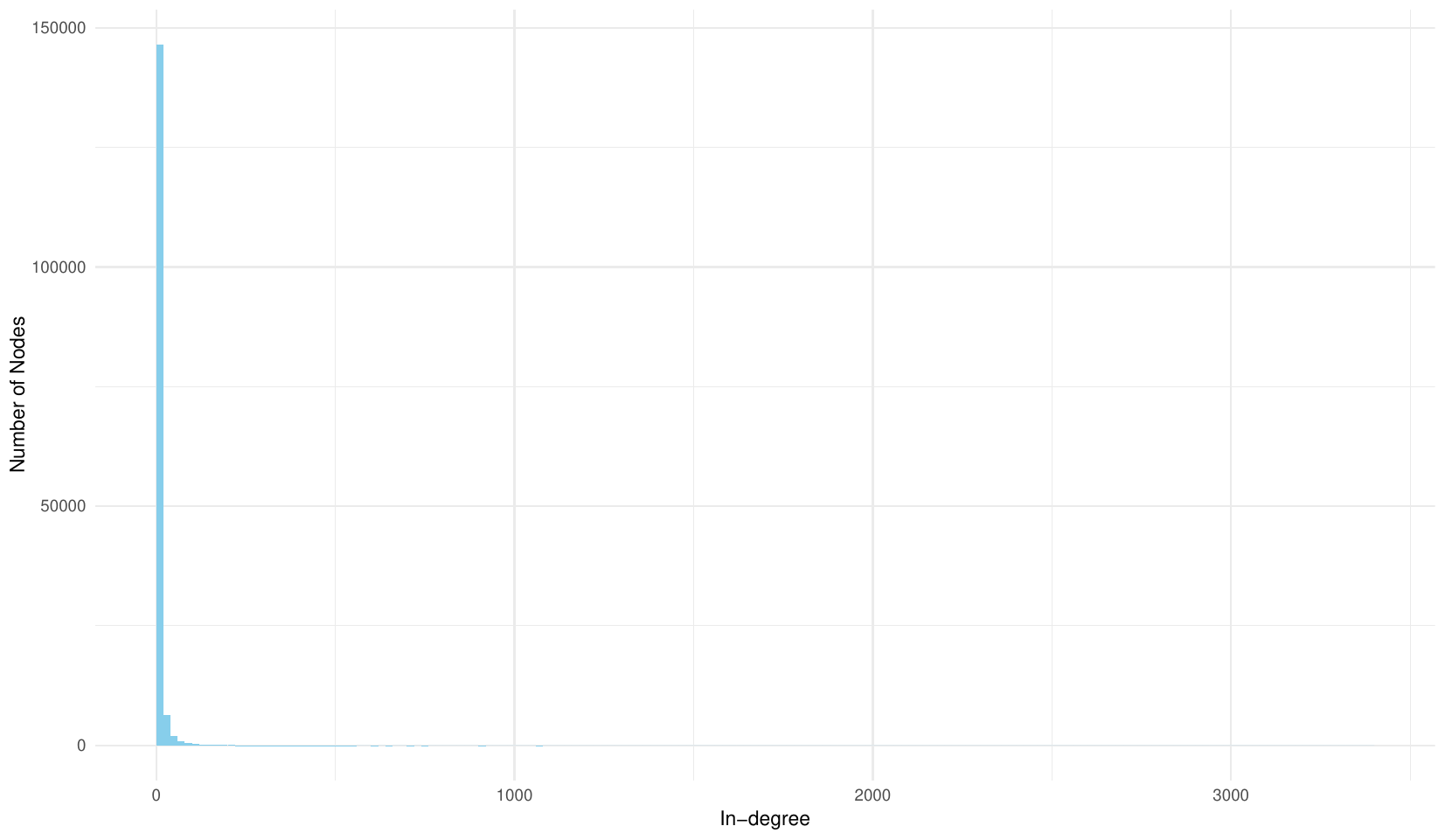}
        \caption{Histogram of in-degree in the citation network.}
        \label{fig:indegreehist}
    \end{subfigure}
    \hfill
    \begin{subfigure}[b]{0.48\textwidth}
        \centering
        \includegraphics[width=\textwidth]{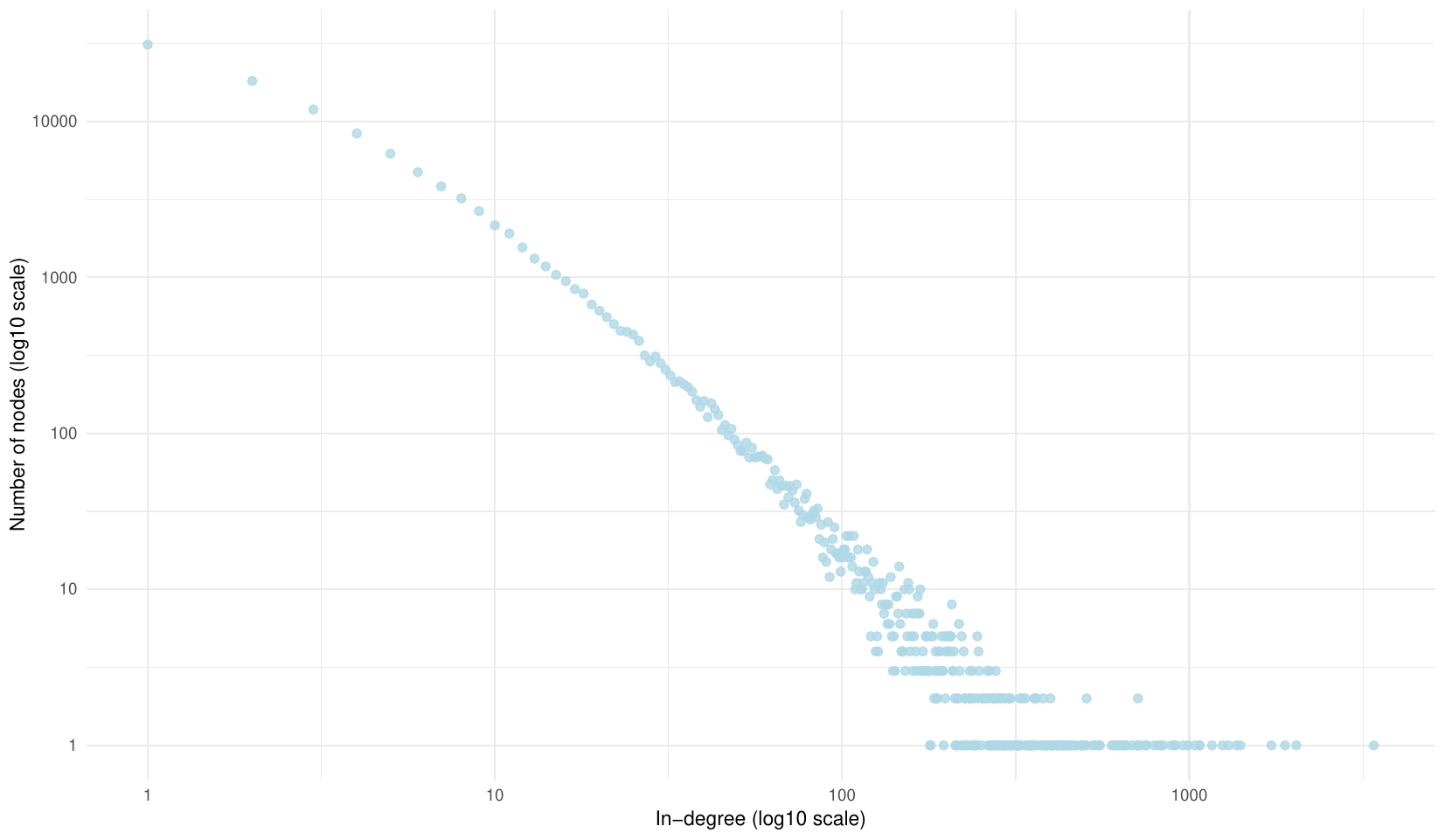}
        \caption{Log–log plot of in-degree in the citation network.}
        \label{fig:log-log}
    \end{subfigure}
    \caption{In-degree distribution of the citation network.}
    \label{fig:indegreedis}
\end{figure}

We next investigate citation patterns through the distribution of received citations in the paper citation network, as measured by nodal in-degree. Nodal in-degree serves as a fundamental indicator of scholarly visibility and impact in citation networks \citep{ZENG20171}. Specifically, the in-degree of article $i$ in the paper citation network is defined as $d_{i}^+ = \sum_{i'\not=i}A_{i'i}$, which represents the total number of citations received by article $i$. Figures \ref{fig:indegreehist} and \ref{fig:log-log} present the in-degree distribution of the paper citation network as a histogram and on a log-log scale. The distribution is highly right-skewed, with the vast majority of articles receiving relatively few citations, while a small number of articles attract a large number of citations. This phenomenon reveals substantial heterogeneity in scholarly impact among publications. For instance, the top 20.13\% of articles account for approximately 80\% of all received citations, highlighting the concentration of scientific influence among a relatively small subset of publications.
On the log–log scale, the distribution exhibits an approximately linear trend over a broad range of values, suggesting heavy-tailed behavior. Such a pattern has been widely documented in empirical studies of citation networks \citep{redner1998popular,traag2025citation}. This indicates that scientific influence is concentrated among a relatively small subset of highly cited publications.

\noindent
{\bf Remark 3.} The in-degree considered in this work differs from citation counts reported by databases such as WoS or Google Scholar. It is defined with respect to the constructed citation network and therefore reflects citations received from papers contained in the dataset. Citations originating from publications outside the selected journals are not included.

\subsection{Derived Network Structures}

\begin{figure}[htbp]
    \centering
    \begin{subfigure}[b]{0.48\textwidth}
        \centering
        \includegraphics[width=\textwidth]{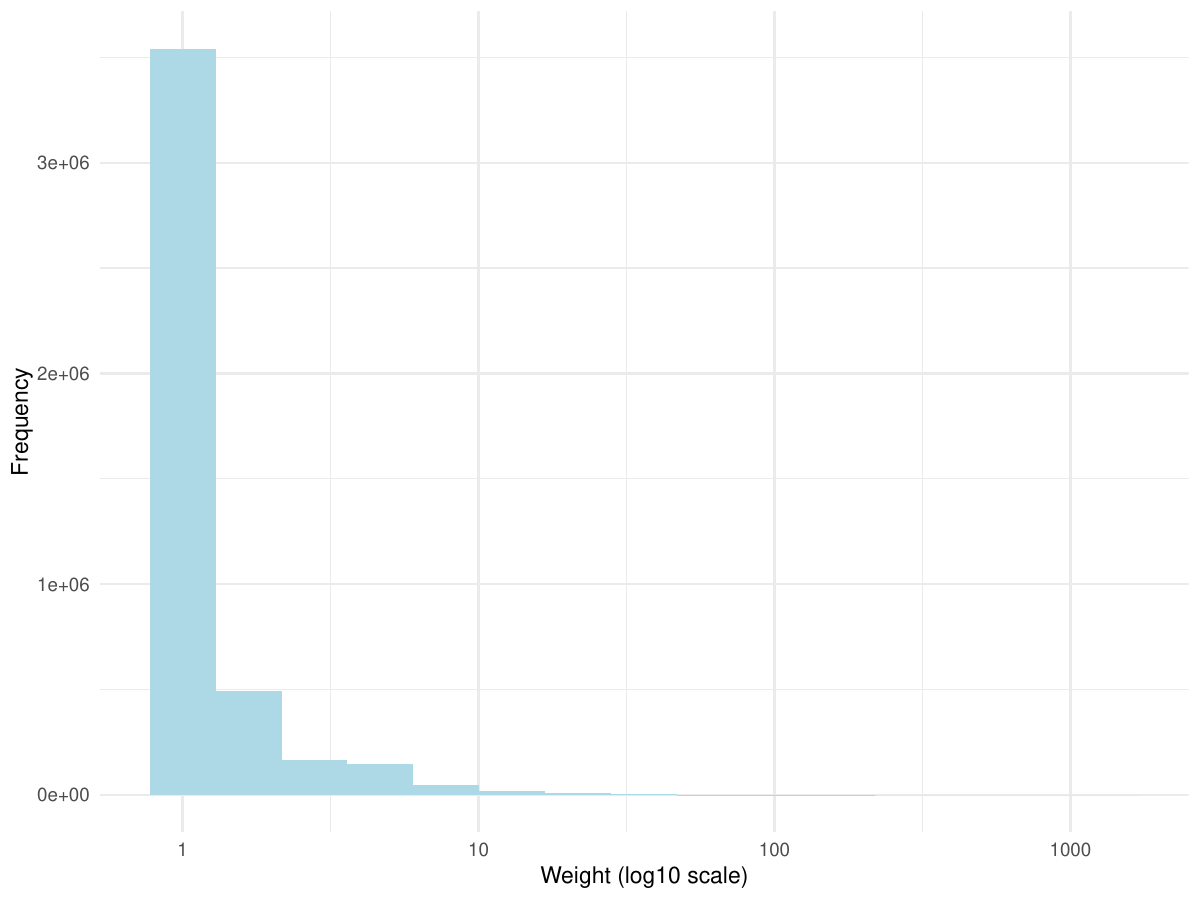}
        \caption{Edge-weight distribution of the co-citation network.}
        \label{fig:cocitation_weight}
    \end{subfigure}
    \hfill
    \begin{subfigure}[b]{0.48\textwidth}
        \centering
        \includegraphics[width=\textwidth]{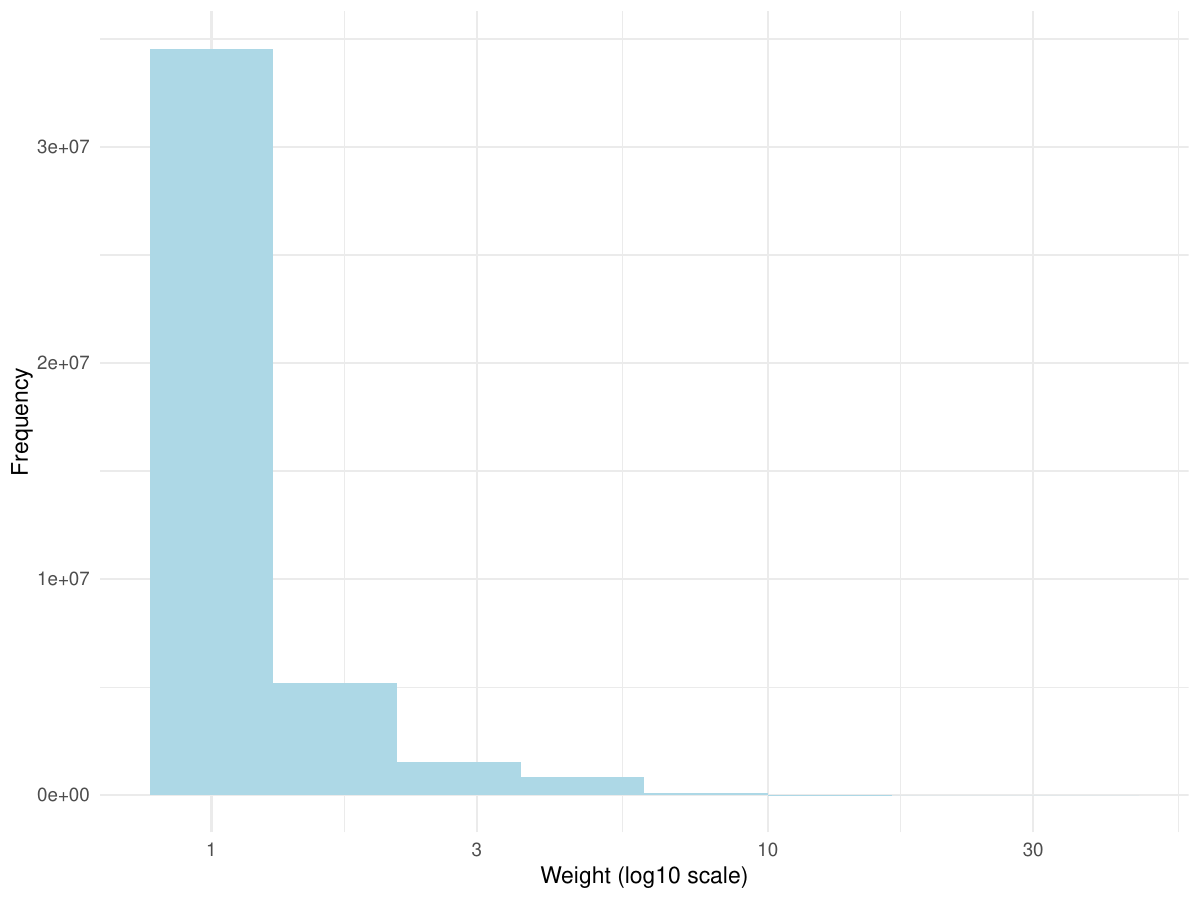}
        \caption{Edge-weight distribution of the bibliographic coupling network.}
        \label{fig:bib_weight}
    \end{subfigure}
    \caption{Edge-weight distributions of the co-citation and bibliographic coupling networks.}
    \label{fig:edgeweight}
\end{figure}

We further examine the structural properties of the co-citation and bibliographic coupling networks. Both networks are undirected and weighted, where edge weights quantify the similarity between pairs of articles through either shared citations or shared references. Figure \ref{fig:edgeweight} presents the edge-weight distributions of the two networks. In both cases, the distributions are highly right-skewed, indicating that most article pairs exhibit only weak similarity, while a relatively small number of pairs form strong connections. For example, 91.03\% of edges in the co-citation network and 94.13\% of edges in the bibliographic coupling network have weights no greater than 2. In contrast, a small fraction of article pairs exhibit strong similarity, with edge weights exceeding 100. Take \cite{tibshirani1996regression} and \cite{fan2001variable} as a concrete example, they are co-cited 1,320 times in the co-citation network. These findings suggest that meaningful relationships are concentrated among a limited subset of articles despite the large number of potential connections.

We further investigate the global connectivity of the two networks through their connected components. The largest connected component contains 98.97\% of all nodes in the co-citation network and 99.31\% in the bibliographic coupling network, indicating that most articles are embedded within a large interconnected structure. At the same time, a number of smaller disconnected components remain, reflecting the existence of specialized research topics and relatively isolated groups of publications. Such a combination of a dominant giant component and many smaller components is commonly observed in large-scale scholarly networks \citep{newman2001scientific}.

\begin{figure}[!ht]
    \centering
    \includegraphics[
    width=0.75\linewidth,
    height=0.75\linewidth,
    keepaspectratio
]{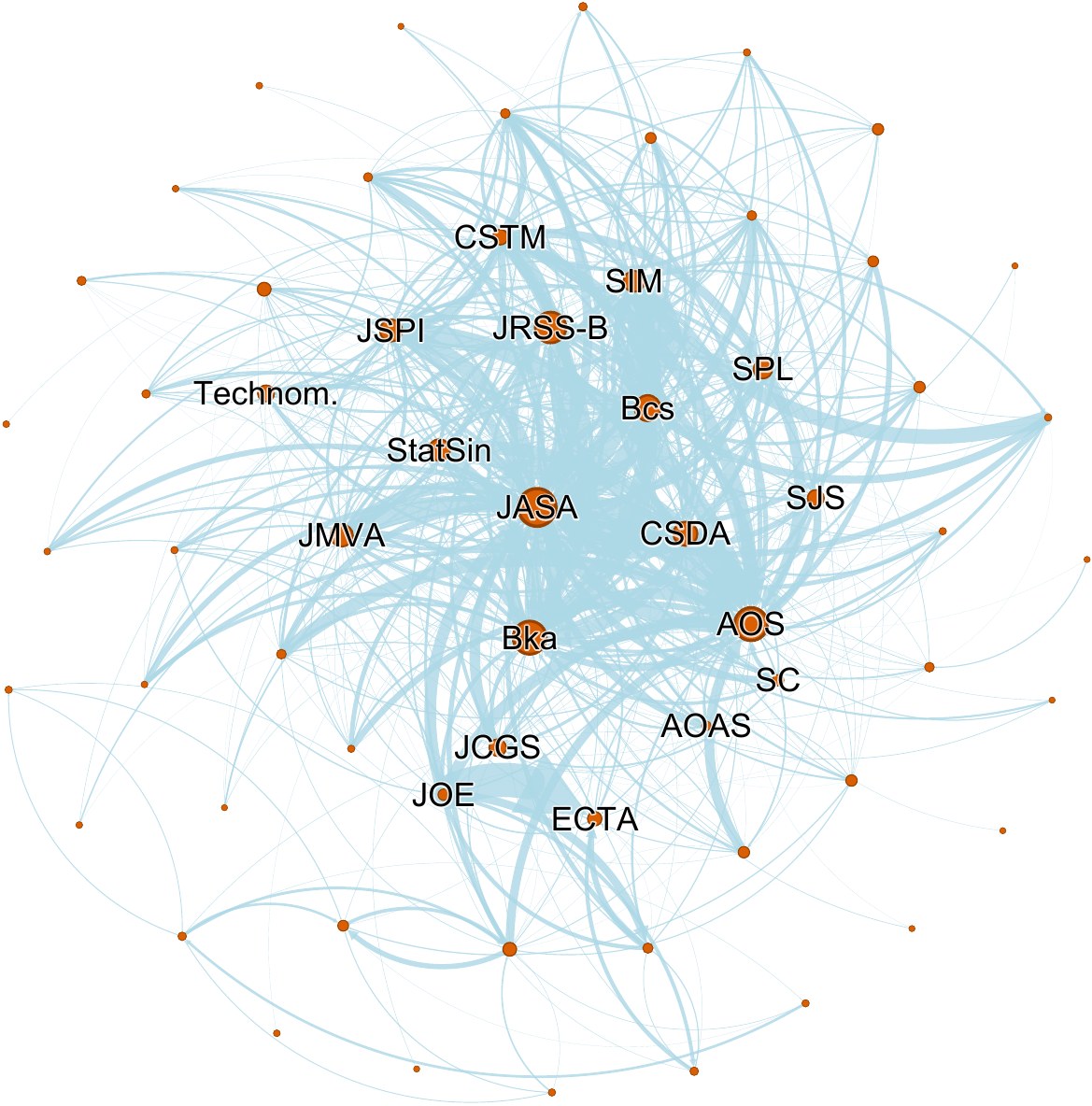}
    \caption{Journal citation network. Only citation links with edge weights greater than or equal to 300 are retained for visualization purposes. Node size is proportional to the in-degree of each journal.}
    \label{fig:journalcite}
\end{figure}

At the journal level, the citation network provides an aggregated view of citation relationships among journals. Figure \ref{fig:journalcite} displays the journal citation network, where only edges with weights of at least 300 are shown for clarity. Compared with the paper citation network, the journal network is considerably denser because each node represents a journal rather than an individual article. The four flagship statistics journals, namely \emph{Journal of the American Statistical Association}, \emph{The Annals of Statistics}, \emph{Biometrika}, and \emph{Journal of the Royal Statistical Society Series B}, are located near the center of the network and maintain citation relationships with many other journals. The figure provides a concise visualization of citation flows among journals in statistics, econometrics, and data mining. The journal citation network may also be used in subsequent studies of journal clustering, hierarchical community detection, and the evolution of research fields.

\subsection{Community Structure}

To illustrate the potential use of the constructed networks, we investigate the community structure of the paper citation network. The purpose is not to develop a new methodology, but rather to demonstrate how the released dataset can support statistical analysis. Among the constructed networks, the paper citation network provides a direct representation of knowledge flows among individual publications and is therefore selected for the community analysis. 

Since the original paper citation network contains a large number of articles with limited citation connections, we first extract a core subgraph to facilitate the subsequent analysis. Specifically, we extract a $q$-core subgraph by retaining articles whose total degree, defined as the sum of the in-degree and out-degree within the induced subgraph, is at least $q$. We use the total degree rather than the in-degree alone because the latter produces an excessively sparse subgraph. Similar preprocessing steps are commonly adopted in the analysis of large-scale citation networks to remove weakly connected nodes while preserving the main connectivity structure \citep{10.1214/16-AOAS977,zhang2023community,zhang2025latent}. In this study, we set $q=13$, which provides a reasonably large connected subgraph while excluding many peripheral nodes. The resulting 13-core contains 25,324 nodes and 354,450 citation links. Community detection is then performed on this subgraph.

We then apply the D-SCORE method \citep{ji2016coauthorship} to identify community structure on the extracted 13-core. D-SCORE is suitable for directed citation networks, as it accounts for asymmetric citation patterns and degree heterogeneity. In this study, it is used as a standard analytical tool rather than a methodological contribution. We consider several candidate values for the number of communities and select the partition that yields the largest modularity (0.433). The resulting network is divided into 9 communities, each representing a group of articles with relatively dense citation relationships. 

\begin{figure}[!ht]
    \centering
    \includegraphics[width=0.9\linewidth]{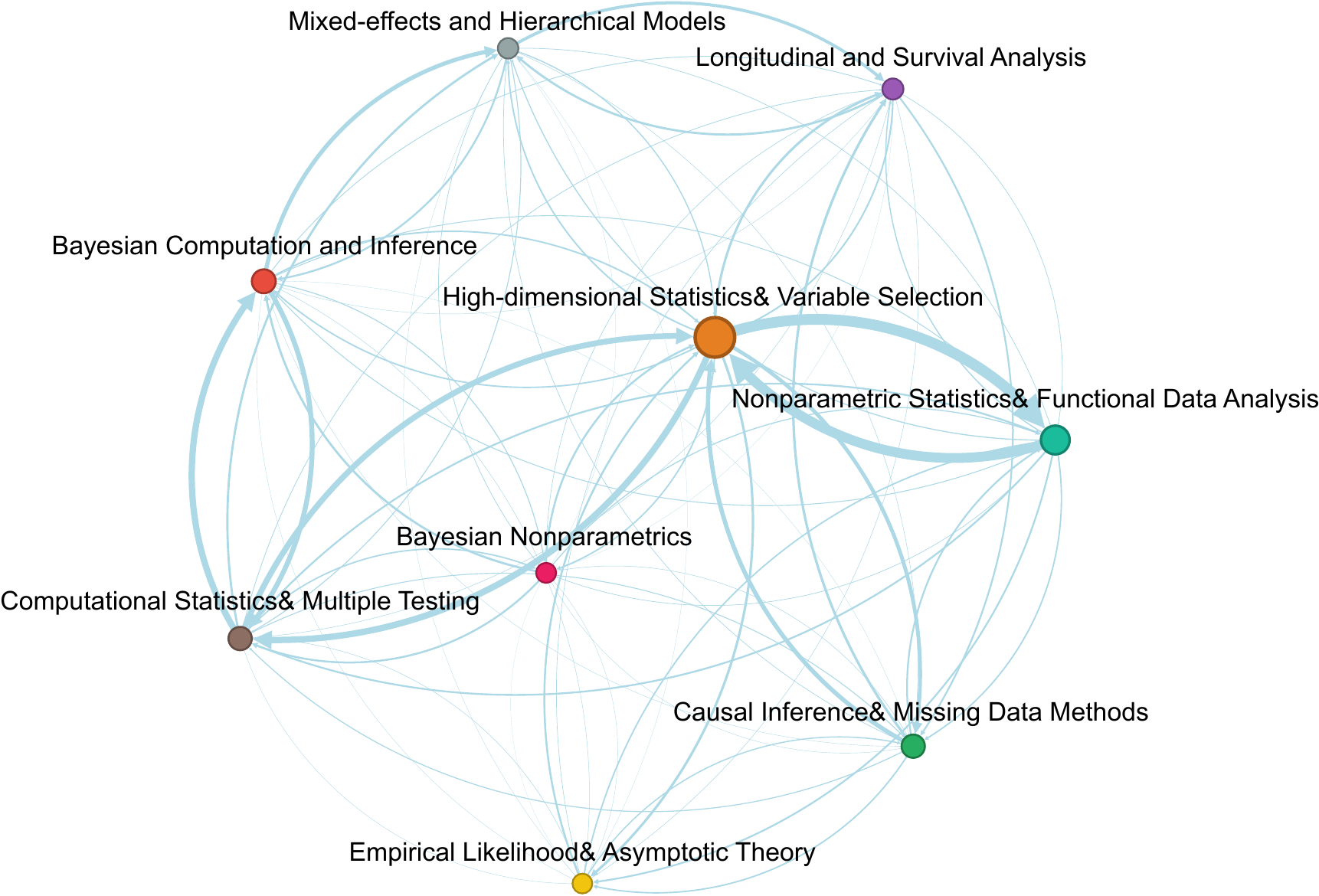}
    \caption{Citation relationships among the detected research communities. Node sizes are proportional to community sizes, and edge widths represent citation frequencies between communities.}
    \label{fig:community}
\end{figure}

To facilitate interpretation, Table~\ref{tab:community_topic} summarizes the detected communities using their representative keywords. The identified communities correspond to several major research areas in statistics and data science, including high-dimensional statistics, nonparametric statistics, Bayesian methods, causal inference, longitudinal analysis, mixed-effects models, and empirical likelihood. The community sizes are highly heterogeneous. High-dimensional statistics and variable selection form the largest community, followed by nonparametric statistics and functional data analysis. Although each community is characterized by distinct topics, several methodological themes, such as Bayesian computation, appear across multiple communities, reflecting the close interactions among different areas of modern statistical research.
Figure~\ref{fig:community} further visualizes the citation relationships among the detected communities. The network shows that the major research areas are closely connected rather than isolated. In particular, high-dimensional statistics and variable selection occupy a central position and maintain strong citation relationships with several neighboring communities, including nonparametric statistics, Bayesian methods, and causal inference. These observations suggest that the citation network captures not only distinct research topics but also their interactions, providing a meaningful representation of the knowledge structure in statistics and data science.

\begin{table}[htbp]
\centering
\caption{Summary of the detected research communities and their representative.}
\label{tab:community_topic}

\small
\renewcommand{\arraystretch}{1.15}
\setlength{\tabcolsep}{5pt}

\begin{tabularx}{\textwidth}{
    >{\centering\arraybackslash}p{1.5cm}
    >{\centering\arraybackslash}p{1.8cm}
    >{\raggedright\arraybackslash}p{4.8cm}
    >{\raggedright\arraybackslash}X
}
\toprule
\textbf{Community} &
\textbf{Size} &
\textbf{Topic} &
\textbf{Representative Keywords} \\
\midrule

C0 & 8,224 &
High-dimensional Statistics \& Variable Selection &
variable selection; LASSO; model selection; sparsity \\

C1 & 4,337 &
Nonparametric Statistics \& Functional Data Analysis &
nonparametric regression; functional data analysis; kernel smoothing; bandwidth selection \\

C2 & 2,593 &
Bayesian Computation and Inference &
Markov chain Monte Carlo; Bayesian inference; EM algorithm; Gibbs sampling \\

C3 & 2,395 &
Computational Statistics \& Multiple Testing &
Markov chain Monte Carlo; false discovery rate; multiple testing; computer experiments \\

C4 & 2,392 &
Causal Inference \& Missing Data Methods &
causal inference; propensity score; missing data; instrumental variables \\

C5 & 1,595 &
Longitudinal and Survival Analysis &
longitudinal data; generalized estimating equations; survival analysis; competing risks \\

C6 & 1,398 &
Mixed-effects and Hierarchical Models &
random effects; longitudinal data; generalized linear mixed model; variance components \\

C7 & 1,242 &
Bayesian Nonparametrics &
Dirichlet process; Bayesian nonparametrics; mixture models; density estimation \\

C8 & 1,148 &
Empirical Likelihood \& Asymptotic Theory &
empirical likelihood; bootstrap; confidence region; asymptotic normality \\

\bottomrule
\end{tabularx}
\end{table}

\section{Discussion}

The constructed dataset provides multiple representations of citation relationships and supports a broad range of statistical analysis. Unlike traditional bibliographic datasets that mainly contain publication records and citation counts, StatCite includes several citation-based networks, including direct citation, co-citation, bibliographic coupling, and journal citation networks. These networks characterize scientific relationships from complementary perspectives: direct citation networks describe knowledge flows among publications, whereas co-citation and bibliographic coupling networks capture intellectual structures and research similarities from retrospective and prospective views, respectively. Such multiple representations provide a more comprehensive description of the organization and evolution of scientific knowledge in statistics and data science \citep{ji2016coauthorship,ji2022co}.

The availability of multiple citation-based networks enables studies from the perspective of multi-layer and heterogeneous networks. Different network layers describe distinct types of relationships among publications and journals, and their joint analysis can provide additional insights beyond a single network representation. For example, researchers can investigate the consistency and differences between citation relationships, intellectual similarity, and journal-level interactions, as well as develop network-based methods for community detection, link prediction, and knowledge structure analysis. Such multi-layer representations have become an important framework for studying complex systems with multiple types of interactions \citep{gao2024citation}.

In addition to network information, StatCite contains rich textual metadata, including titles, abstracts, and keywords. The combination of textual information and citation networks provides opportunities for text-enhanced scientometric analysis. Recent studies have emphasized the potential of integrating text mining methods with science mapping approaches to improve topic identification and knowledge discovery in large-scale bibliographic analysis \citep{chen2023text,https://doi.org/10.1002/sta4.545,annurev-statistics-040522-022138}. The textual and network information available in StatCite can support studies on topic evolution, research novelty, semantic similarity, knowledge transfer, and recommendation systems. Furthermore, the dataset may facilitate emerging applications involving large language models, such as automated literature analysis, research trend identification, and intelligent academic information retrieval \citep{Gao02012026}.

Future development of StatCite will focus on extending and maintaining the coverage of the dataset. As the literature in statistics and data science continues to expand, future updates can incorporate newly published articles and additional journals to provide a more comprehensive representation of the research landscape. In addition, future versions may include further types of scholarly relationships, such as author collaboration networks, institutional networks, and topic networks, to provide additional perspectives on the structure of scientific activities. Improvements in metadata coverage and data quality control can also further enhance the utility of StatCite.

% Bibliography using author-year style
\bibliographystyle{plainnat} % Author-year style
\bibliography{references} % Your BibTeX file

@article{Gao02012026,
author = {Tianchen Gao and Jiashun Jin and Zheng Tracy Ke and Gabriel Moryoussef},
title = {A Comparison of DeepSeek and other LLMs},
journal = {The American Statistician},
volume = {80},
number = {1},
pages = {164--176},
year = {2026},
publisher = {Taylor \& Francis},
doi = {10.1080/00031305.2025.2611010}
}

@article{gao2024citation,
  author  = {Gao, Tianchen and Liu, Jingyuan and Pan, Rui and Wang, Hansheng},
  title   = {Citation Counts Prediction of Statistical Publications Based on
             Multi-Layer Academic Networks via Neural Network Model},
  journal = {Expert Systems with Applications},
  year    = {2024},
  volume  = {238},
  pages   = {121634},
  doi     = {10.1016/j.eswa.2023.121634}
}

@article{annurev-statistics-040522-022138,
   author = {Ke, Zheng Tracy and Ji, Pengsheng and Jin, Jiashun and Li, Wanshan},
   title = {Recent Advances in Text Analysis}, 
   journal= {Annual Review of Statistics and Its Application},
   year = {2024},
   volume = {11},
   pages   = {347--372}
  }

@article{https://doi.org/10.1002/sta4.545,
author = {Ke, Zheng Tracy and Jin, Jiashun},
title = {Special invited paper: The SCORE normalization, especially for heterogeneous network and text data},
journal = {Stat},
volume = {12},
number = {1},
pages = {e545},
doi = {https://doi.org/10.1002/sta4.545},
url = {https://onlinelibrary.wiley.com/doi/abs/10.1002/sta4.545},
eprint = {https://onlinelibrary.wiley.com/doi/pdf/10.1002/sta4.545},
year = {2023}
}

@article{chen2023text,
  title={When text mining meets science mapping in the bibliometric analysis: A review and future opportunities},
  author={Chen, Haojing and Tsang, Yung P and Wu, Chun H},
  journal={International Journal of Engineering Business Management},
  volume={15},
  pages={18479790231222349},
  year={2023},
  publisher={SAGE Publications Sage UK: London, England}
}

@article{zhang2025latent,
  title={A latent space model for weighted keyword Co-occurrence networks with applications in knowledge discovery in statistics},
  author={Zhang, Yan and Pan, Rui and Zhu, Xuening and Fang, Kuangnan and Wang, Hansheng},
  journal={Journal of Computational and Graphical Statistics},
  volume={34},
  number={3},
  pages={779--794},
  year={2025},
  publisher={Taylor \& Francis}
}

@article{zhang2023community,
  title={Community detection in attributed collaboration network for statisticians},
  author={Zhang, Yan and Pan, Rui and Wang, Hansheng and Su, Haibo},
  journal={Stat},
  volume={12},
  number={1},
  pages={e507},
  year={2023},
  publisher={Wiley Online Library}
}

@article{10.1214/16-AOAS977,
author = {Song Wang and Karl Rohe},
title = {{Discussion of “Coauthorship and citation networks for statisticians”}},
volume = {10},
journal = {The Annals of Applied Statistics},
number = {4},
publisher = {Institute of Mathematical Statistics},
pages = {1820 -- 1826},
year = {2016},
doi = {10.1214/16-AOAS977},
URL = {https://doi.org/10.1214/16-AOAS977}
}

@incollection{traag2025citation,
  title={Citation models and research evaluation},
  author={Traag, Vincent A.},
  editor={Yasseri, Taha},
  booktitle={Handbook of Computational Social Science},
  year={2025},
  publisher={Edward Elgar Publishing}
}

@article{redner1998popular,
  title={How popular is your paper? An empirical study of the citation distribution},
  author={Redner, Sidney},
  journal={The European Physical Journal B-Condensed Matter and Complex Systems},
  volume={4},
  number={2},
  pages={131--134},
  year={1998},
  publisher={Springer}
}

@article{ZENG20171,
title = {The science of science: From the perspective of complex systems},
journal = {Physics Reports},
volume = {714-715},
pages = {1-73},
year = {2017},
note = {The Science of Science: From the Perspective of Complex Systems},
issn = {0370-1573},
doi = {https://doi.org/10.1016/j.physrep.2017.10.001},
url = {https://www.sciencedirect.com/science/article/pii/S0370157317303289},
author = {An Zeng and Zhesi Shen and Jianlin Zhou and Jinshan Wu and Ying Fan and Yougui Wang and H. Eugene Stanley}
}

@article{He12012026,
author = {Chenxuan He and Feifei Wang and Liping Zhu},
title = {Emerging Knowledge Trend in Statistical Research: A Content-Based Analysis Using Covariate-Assisted Dynamic Topic Model},
journal = {Journal of the American Statistical Association},
volume = {0},
number = {0},
pages = {1--23},
year = {2026},
publisher = {Taylor \& Francis}
}

@article{ji2022co,
  title={Co-citation and co-authorship networks of statisticians},
  author={Ji, Pengsheng and Jin, Jiashun and Ke, Zheng Tracy and Li, Wanshan},
  journal={Journal of Business \& Economic Statistics},
  volume={40},
  number={2},
  pages={469--485},
  year={2022},
  publisher={Taylor \& Francis}
}

@article{ji2016coauthorship,
author = {Pengsheng Ji and Jiashun Jin},
title = {{Coauthorship and citation networks for statisticians}},
volume = {10},
journal = {The Annals of Applied Statistics},
number = {4},
publisher = {Institute of Mathematical Statistics},
pages = {1779 -- 1812},
year = {2016},
doi = {10.1214/15-AOAS896},
URL = {https://doi.org/10.1214/15-AOAS896}
}

@article{hutchins2019nih,
  title={The NIH Open Citation Collection: A public access, broad coverage resource},
  author={Hutchins, B Ian and Baker, Kirk L and Davis, Matthew T and Diwersy, Mario A and Haque, Ehsanul and Harriman, Robert M and Hoppe, Travis A and Leicht, Stephen A and Meyer, Payam and Santangelo, George M},
  journal={PLoS Biology},
  volume={17},
  number={10},
  pages={e3000385},
  year={2019},
  publisher={Public Library of Science San Francisco, CA USA}
}

@inproceedings{leskovec2005graphs,
  title={Graphs over time: densification laws, shrinking diameters and possible explanations},
  author={Leskovec, Jure and Kleinberg, Jon and Faloutsos, Christos},
  booktitle={Proceedings of the Eleventh ACM SIGKDD International Conference on Knowledge Discovery in Data Mining},
  pages={177--187},
  year={2005}
}

@article{bornmann2025citation,
  title={Citation accuracy, citation noise, and citation bias: A foundation of citation analysis},
  author={Bornmann, Lutz and Leibel, Christian},
  journal={arXiv preprint arXiv:2508.12735},
  year={2025}
}

@article{price1965networks,
  title={Networks of scientific papers: The pattern of bibliographic references indicates the nature of the scientific research front.},
  author={Price, Derek J De Solla},
  journal={Science},
  volume={149},
  number={3683},
  pages={510--515},
  year={1965},
  publisher={American Association for the Advancement of Science}
}

@article{krishen2021broad,
  title={A broad overview of interactive digital marketing: A bibliometric network analysis},
  author={Krishen, Anjala S and Dwivedi, Yogesh K and Bindu, N and Kumar, K Satheesh},
  journal={Journal of Business Research},
  volume={131},
  pages={183--195},
  year={2021},
  publisher={Elsevier}
}

@article{liu2025academic,
  title={Academic Literature Recommendation in Large-scale Citation Networks Enhanced by Large Language Models},
  author={Liu, Kun and Zhang, Yan and Pan, Rui and Gao, Tianchen and Wang, Hansheng},
  journal={Scientometrics},
  volume={130},
  number={9},
  pages={5143--5169},
  year={2025}
}

@article{gao2021community,
  title={Community detection for statistical citation network by D-SCORE},
  author={Gao, Tianchen and Pan, Rui and Wang, Siyu and Yang, Yuehan and Zhang, Yan},
  journal={Statistics and Its Interface},
  volume={14},
  number={3},
  pages={279--294},
  year={2021},
  publisher={International Press of Boston}
}

@article{gao2024community,
  title={Community detection in temporal citation network via a tensor-based approach},
  author={Gao, Tianchen and Pan, Rui and Zhang, Junfei and Wang, Hansheng},
  journal={Statistics and Its Interface},
  volume={17},
  number={2},
  pages={145--158},
  year={2024},
  publisher={International Press of Boston}
}

@article{feng2024citation,
  title={Citation network analysis of retractions in molecular biology field},
  author={Feng, Sida and Feng, Lingzi and Han, Fang and Zhang, Ye and Ren, Yanqing and Wang, Lixue and Yuan, Junpeng},
  journal={Scientometrics},
  volume={129},
  number={8},
  pages={4795--4817},
  year={2024},
  publisher={Springer}
}

@article{teich2022citation,
  title={Citation inequity and gendered citation practices in contemporary physics},
  author={Teich, Erin G and Kim, Jason Z and Lynn, Christopher W and Simon, Samantha C and Klishin, Andrei A and Szymula, Karol P and Srivastava, Pragya and Bassett, Lee C and Zurn, Perry and Dworkin, Jordan D and others},
  journal={Nature Physics},
  volume={18},
  number={10},
  pages={1161--1170},
  year={2022},
  publisher={Nature Publishing Group UK London}
}

@article{varin2016statistical,
  title={Statistical modelling of citation exchange between statistics journals},
  author={Varin, Cristiano and Cattelan, Manuela and Firth, David},
  journal={Journal of the Royal Statistical Society Series A: Statistics in Society},
  volume={179},
  number={1},
  pages={1--63},
  year={2016},
  publisher={Oxford University Press}
}

@article{gao2023large,
  title={Large-scale multi-layer academic networks derived from statistical publications},
  author={Gao, Tianchen and Zhang, Yan and Pan, Rui and Wang, Hansheng},
  journal={arXiv preprint arXiv:2308.11287},
  year={2023}
}

@article{fortunato2018science,
  title={Science of science},
  author={Fortunato, Santo and Bergstrom, Carl T and B{\"o}rner, Katy and Evans, James A and Helbing, Dirk and Milojevi{\'c}, Sta{\v{s}}a and Petersen, Alexander M and Radicchi, Filippo and Sinatra, Roberta and Uzzi, Brian and others},
  journal={Science},
  volume={359},
  number={6379},
  pages={eaao0185},
  year={2018},
  publisher={American Association for the Advancement of Science}
}

@article{mejia2021exploring,
  title={Exploring topics in bibliometric research through citation networks and semantic analysis},
  author={Mejia, Cristian and Wu, Mengjia and Zhang, Yi and Kajikawa, Yuya},
  journal={Frontiers in Research Metrics and Analytics},
  volume={6},
  pages={742311},
  year={2021},
  publisher={Frontiers Media SA}
}

@article{mingers2015review,
  title={A review of theory and practice in scientometrics},
  author={Mingers, John and Leydesdorff, Loet},
  journal={European Journal of Operational Research},
  volume={246},
  number={1},
  pages={1--19},
  year={2015},
  publisher={Elsevier}
}

@article{chen2010structure,
  title={The structure and dynamics of cocitation clusters: A multiple-perspective cocitation analysis},
  author={Chen, Chaomei and Ibekwe-SanJuan, Fidelia and Hou, Jianhua},
  journal={Journal of the American Society for information Science and Technology},
  volume={61},
  number={7},
  pages={1386--1409},
  year={2010},
  publisher={Wiley Online Library}
}

@article{kessler1963bibliographic,
  title={Bibliographic coupling between scientific papers},
  author={Kessler, Maxwell Mirton},
  journal={American Documentation},
  volume={14},
  number={1},
  pages={10--25},
  year={1963},
  publisher={Wiley Online Library}
}

@article{tibshirani1996regression,
  title={Regression shrinkage and selection via the lasso},
  author={Tibshirani, Robert},
  journal={Journal of the Royal Statistical Society Series B: Statistical Methodology},
  volume={58},
  number={1},
  pages={267--288},
  year={1996},
  publisher={Oxford University Press}
}

@article{fan2001variable,
  title={Variable selection via nonconcave penalized likelihood and its oracle properties},
  author={Fan, Jianqing and Li, Runze},
  journal={Journal of the American statistical Association},
  volume={96},
  number={456},
  pages={1348--1360},
  year={2001},
  publisher={Taylor \& Francis}
}

@article{newman2001scientific,
  title={Scientific collaboration networks. I. Network construction and fundamental results},
  author={Newman, Mark EJ},
  journal={Physical Review E},
  volume={64},
  number={1},
  pages={016131},
  year={2001},
  publisher={APS}
}

\end{document}